\documentclass[reprint,footinbib,amsmath,amssymb,aps,prl,floatfix,superscriptaddress,a4paper]{revtex4-2}
\usepackage{graphicx}
\usepackage{dcolumn}
\usepackage{bm}
\usepackage[colorlinks=true,allcolors = blue]{hyperref}
\usepackage{newtxtext,newtxmath}
\usepackage{siunitx}
\usepackage{microtype}
\usepackage{selnolig}

\newcommand{\leibnizd}[1]{\mathrm{d}{#1}}
\newcommand{\load}{N}
\newcommand{\drag}{F}
\newcommand{\gammadot}{{\dot\gamma}}
\newcommand{\order}[1]{\mathcal{O}({#1})}
\newcommand{\Eq}[1]{Eq.~\eqref{#1}}
\newcommand{\Fig}[1]{Fig.~\ref{#1}}
\newcommand{\partFig}[2]{Fig.~\hyperref[#1]{\ref*{#1}#2}}
\newcommand{\ecfp}{Edinburgh Complex Fluids Partnership and School of Physics and Astronomy, %
  The~University~of~Edinburgh, James Clerk Maxwell Building, Peter Guthrie Tait Road, %
  Edinburgh EH9 3FD, United Kingdom}
\newcommand{\Refcite}[1]{Ref.~[\onlinecite{#1}]} 
\newcommand{\Refscite}[1]{Refs.~[\onlinecite{#1}]} 
\newcommand{\naive}{{na\"\i{}ve}}
\newcommand{\latin}[1]{{\itshape #1}}
\newcommand{\eg}{\latin{e.g.}}
\newcommand{\ie}{\latin{i.e.}}
\newcommand{\cf}{\latin{cf.}}
\newcommand{\interalia}{\latin{inter alia}}

\begin{document}

\title{Anomalous Scaling for Hydrodynamic Lubrication of Conformal Surfaces}

\author{James A. Richards}
\email{james.a.richards@ed.ac.uk}
\affiliation{\ecfp}
\author{Patrick B. Warren}
\email{patrick.warren@stfc.ac.uk}
\affiliation{The Hartree Centre, STFC Daresbury Laboratory, Warrington WA4 4AD, United Kingdom}

\author{Daniel J. M. Hodgson}
\affiliation{\ecfp}
\author{Alex Lips}
\affiliation{\ecfp}
\author{Wilson C. K. Poon}
\email{w.poon@ed.ac.uk}
\affiliation{\ecfp}


\begin{abstract}
  The hydrodynamic regime of the Stribeck curve giving the friction coefficient $\mu$ as a function of the dimensionless relative sliding speed (the Sommerfeld number, $S$) of two contacting non-conformal surfaces is usually considered trivial, with $\mu \sim S$. We predict that for conformal surfaces contacting over large areas, a combination of independent length scales gives rise to a universal power-law with a non-trivial exponent, $\mu\sim S^{2/3}$, for a thick lubrication film.  Deviations as the film thins (decreasing $S$) may superficially resemble the onset of elastohydrodynamic lubrication, but are due to a crossover between hydrodynamic regimes.  Our experiments as well as recent measurements of chocolate lubrication confirm these predictions.
\end{abstract}
\maketitle

Controlling friction between sliding surfaces is important across multiple fields~\cite{hamrock2004fundamentals}. For example, friction losses in bearings account for a third of a car's fuel use and 23\% of global energy production~\cite{holmberg2019impact}. Tribological properties determine the sensory feel of topical products such as skin creams~\cite{adams2007friction,akay2012measurement}. Inter-particle friction  is implicated in suspension rheology~\cite{lin2015hydrodynamic}.
Lubricants, from mineral oils to the synovial fluid in our joints,  reduce wear 
and frictional losses.  The lubricant viscosity, $\eta$, is one determinant of the friction coefficient, $\mu = \drag/\load$, the drag force ($\drag$) to load ($\load$) ratio; but it also depends on the relative sliding speed, $U$, of the surfaces, their geometry, and the load. Following Stribeck, G\"umbel and Hersey independently proposed a  dimensionless parameter to rationalise these dependencies~\cite{stribeck1902abc, gumbel1914problem, hersey1914laws}.  For a bearing of width $L$, this parameter is $S = \eta U\!L/\load$ (often called the Sommerfeld number).

By the 1920s a canonical view had emerged~\cite{wilson1922mechanism} that the `Stribeck curve', $\mu(S)$, generally displays a minimum, which is usually taken to correspond to the transition from hydrodynamic lubrication (HL) under low loads (high $S$), through elasto-hydrodynamic lubrication (EHL), to boundary lubrication (BL) at high loads (low $S$), with the transition determined by the microscopic asperity length scale.  Deemed understood after the early twentieth century~\cite{sommerfeld1904hydrodynamischen,rayleigh1918notes,kapitza1955hydrodynamic}, the HL regime is usually dismissed: pedagogical discussions often claim casually that $\mu \sim S$ in this regime (\eg,  Fig.~2 in \Refcite{woydt2010history}), with no supporting data. Rather, in engineering tribology the focus shifted to the small-$S$, heavily-loaded $\mathrm{EHL}\to\mathrm{BL}$ transition, where the minimum in $\mu(S)$ gives least wear and dissipation~\cite{hamrock2004fundamentals,stachowiak2013engineering}. The physics in these regimes is complex, and involves coupling solid and fluid mechanics~\cite{fielding2023model} as well as lubricant molecular properties~\cite{stachowiak2013engineering}.

\begin{figure}[b]
    \centering
    \includegraphics{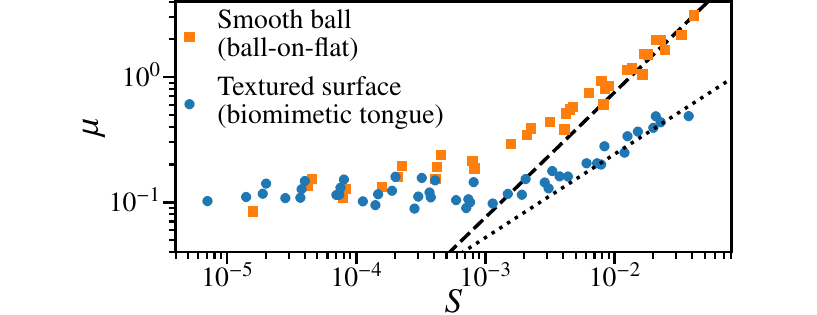}
    \caption{Stribeck curve, friction coefficient ($\mu$) as a function of non-dimensionalised sliding speed (Sommerfeld number, $S$) for various molten chocolate samples in different geometries. Symbols: (orange) squares, ball-on-flat ($R = \SI{6.3}{\milli\metre}$ and $\load = \SI{0.01}{\newton}$) with dashed line, $\mu \sim S$; (blue) circles, textured-surface-on-flat with dotted line, $\mu \sim S^{2/3}$. Replotted from \Refcite{soltanahmadi_insights_2023}; see \Refcite{PRE} for details.}
    \label{fig:tongue}
\end{figure}

Recently, though, the high-$S$, lightly-loaded (or high lubricant viscosity) regime has received renewed attention because of its relevance for human sensory perception, such as oral `mouth-feel'~\cite{Pradal2016}. In particular, a recent study of the lubrication behaviour of molten chocolate~\cite{soltanahmadi_insights_2023} shows data (reproduced in \Fig{fig:tongue}) for a ball-on-flat contact where $\mu(S)$ does indeed appear to tend towards $\mu \sim S$ at high $S$. Significantly, however, for a textured bio-mimetic tongue surface, a different high-$S$ behaviour is found, with a clearly weaker dependence on $S$. 

Here, we show by experiment and theory that in bearing geometries characterised by \emph{two} length scales, a macroscopic bearing dimension and a mesoscopic surface profile length scale, we generally expect $\mu \sim S^{2/3}$ in the high $S$, large-gap, HL limit. The two length scales set a cross-over $S^*$, as the lubrication film thins, below which deviations from $S^{2/3}$ scaling can mimic the well-known EHL upturn but are entirely due to hydrodynamics.  We argue, \interalia, that this explains the bio-mimetic tongue data in \Fig{fig:tongue}.

\begin{figure}
  \centering
  \includegraphics{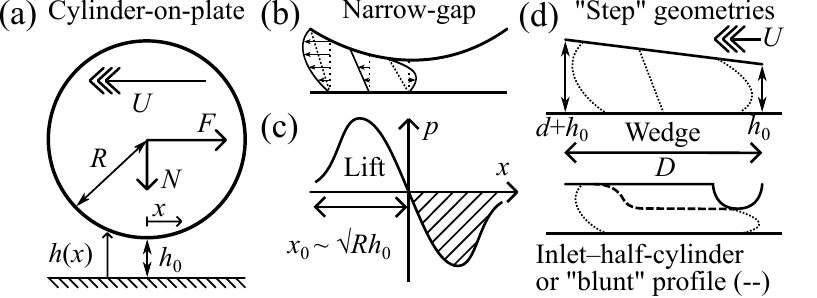}
  \caption{Lubrication geometries. (a)~Cylinder-on-flat: gap, $h(x)$; radius, $R$; and $h_0 = \min(h)$. Conditions: load, $N$; sliding velocity, $U$; and, drag, $F$. Long-bearing into plane, $L\gg x_0$. (b)~Narrow-gap, with Couette (dashed line) and Poiseuille flow (arrows). (c)~Resulting pressure, $p(x)$. Hatching, $p<0$ neglected with half-Sommerfeld approximation. (d)~Conformal contacts with step, $d$; and length, $D$: upper, wedge; lower, inlet--half-cylinder (solid) and step (dashed).}
  \label{fig:ball}
\end{figure}

To introduce our theoretical framework, consider first a canonical \emph{non-conformal} contact comprising a cylinder of radius $R$ sliding against a flat (\partFig{fig:ball}{a}).  Here, the gap is $h \approx h_0 + x^2\!/2R$ with $h_0 \ll R$ the minimum gap height and $x$ the distance from this point. There is a region $x_0\sim\sqrt{Rh_0}$ (\partFig{fig:ball}{b}) in which the gap is $\order{h_0}$, outside of which pressures and stresses are negligibly small. With a characteristic shear rate $\gammadot \sim U/h_0$, the frictional drag force on a cylinder of length $L$ due to Couette flow  (\partFig{fig:ball}{b}, dashed lines) is $\drag \sim \eta \gammadot Lx_0\sim \eta U\!L\sqrt{R/h_0}$.
To conserve volume for incompressible fluids, an additional, compensating Poiseuille flow is needed (\partFig{fig:ball}{b}, solid lines).  The associated `Reynolds lubrication pressure' (\partFig{fig:ball}{c}) generates the load-bearing normal force.  In the `long-bearing' limit ($L \gg x_0$), this compensating Poiseuille flow develops parallel to the Couette flow and, in the cylinder-on-flat case, of a similar magnitude~\cite{warren2016sliding}. The corresponding pressure $p$ emerges from the Hagen-Poiseuille expression, $U\sim (h_0^2/\eta)\times p/x_0$\,; together with the area $\sim Lx_0$ this sets the normal force $\load \sim pLx_0\sim \eta U\! Lx_0^2/h_0^2\sim\eta U\!LR/h_0$.  

This problem is symmetric about $x=0$, so that equal but opposite pressures should be created in the converging and diverging regions (\partFig{fig:ball}{c}). We appeal to a widely used `half-Sommerfeld' boundary condition~\cite{hamrock2004fundamentals} and set the negative pressure in the diverging region to zero. This can be justified when, \eg, the maximum pressure is greater than the difference between the (atmospheric) inlet pressure and the lubricant vapour pressure and cavitation occurs.

Since $h_0$ adjusts to support the load, the friction coefficient $\mu=\drag/\load \sim \sqrt{h_0/R}$ depends on $S=\eta UL/N$. For the  cylinder on flat, one finds $\mu \sim S^{1/2}$ with an $\order{1}$ numerical prefactor and no further dependence on $R$ or lubricant properties.  A similar analysis for a sphere gives $\mu \sim S$~\cite{warren2016sliding}.  These scaling laws apply for non-conformal contacts for all $h_0$ beyond contacting asperities. They occasion no surprise, and reflect the spatial dimension.  This simplicity is traceable to the the fact that the extent of the narrow-gap region is $x_0 \sim \sqrt{Rh_0}$. Thus, the problem is specified by one length scale, $R$, and the magnitude of the induced Poiseuille flow is always $\Delta U \sim U$.

In contrast, and forming part of the historic foundation of tribology, conformal surfaces allow close contact over a wide area~\cite{desplanques2015amontons}. For soft surfaces, such as skin or ceramic green bodies, bulk deformation brings surfaces into broad close approach. At first sight, there are no obvious length scales in a flat-on-flat contact corresponding to $R$ for the sphere or cylinder.  However, large-area surfaces typically show both microscale roughness and mesoscale non-flatness. Studies of artificial `textured' conformal contacts~\cite{gropper2016hydrodynamic} suggest that a general (macroscopically) flat surface can be modelled as the sum of many elementary `texture cells', each of which is a form of slider bearing. Common slider bearing geometries include the pedagogical examples of a (Rayleigh) step~\cite{rayleigh1918notes} and a wedge, to which we add an inlet--half-cylinder, \partFig{fig:ball}{d}. The HL problem in each case can be reduced to quadratures, as detailed in a companion paper \cite{PRE}. Here, we extend our scaling analysis to identify the key generic features. 

The key idea is that a textured surface is characterised by \emph{two} length scales: a `step height' $d$ and `step length' $D$. To conserve volume and balance the changing Couette flow as the gap narrows from $h_0 +d$ to $h_0$, a Poiseuille flow of order $\Delta U \sim Ud/[d+\order{h_0}]$ is required (assuming a `long bearing' limit $L\gg D$\,; see below for short bearings). At modest gaps ($h_0\alt d$) one has $\Delta U\sim U$, as in the case of non-conformal contacts. However, in the large-gap limit ($h_0 \gg d$), $\Delta U \sim U d/h_0 \ll U$. Hagen-Poiseuille, with $D$ replacing $x_0$, now gives $\Delta U\sim (h_0^2/\eta)\times p/D$, and a lift force $N \sim pLD \sim\eta LD^2\Delta U/h_0^2$\, hence the Sommerfeld number $S=\eta U\!L/N \sim h_0^3/D^2d$ for $h_0\gg d$. The Couette flow generates a drag force $\drag \sim \eta U\!L D/h_0$ for $h_0 \gg d$, so that $\mu=F/N \sim h_0^2/Dd$ for $h_0\gg d$.  Eliminating $h_0$ parametrically between $\mu$ and $S$ yields $\mu\sim S^{2/3}$ for $S\gg S^*$, where $S^*$ corresponds to $h_0\approx d$.

The replacement of the power-law $\mu\sim S^{1/2}$ (expected on dimensional grounds) by $\mu\sim S^{2/3}$ is an example of the failure of a `regularity assumption' ~\cite{goldenfeld1992lectures}, which in this problem amounts to the `\naive' assertion that $\Delta U \sim U$ for the compensating Poiseuille flow for all gap sizes. Whilst this is correct for the non-conformal cases of the cylinder and the sphere, and for moderate gaps ($h_0 \sim d$) in conformal contacts, it misses a dimensionless factor of $d/h_0$ in the wide-gap ($h_0 \gg d$) limit.

These results are confirmed by an analysis of the rigorously-derived expressions in lubrication theory~\cite{PRE}, and also hold for the so-called  DuBois-Ocvirk `short-bearing' approximation~\cite{dubois1953analytical}, which will be relevant for our experiments.  In this latter limit, volume conservation of the lubricant in the gap occurs through side leakage and the induced Poiseuille flow is \emph{perpendicular} to the Couette flow.  This appears quite different to the `long bearing' case: indeed the lift force is reduced, with side leakage occurring over a wider width, $D\gg L$, and travelling a shorter length, $L\ll D$. However, the same anomalous exponent still arises in the large-gap HL regime~\cite{PRE}, with pre-factors modified by $L^2/D^2$:
\begin{equation}
    \mu \sim \frac{Dd}{L^2} \Bigl (\frac{h_0}{d} \Bigr)^2,~S \sim
    \frac{d^2}{L^2} \Bigl( \frac{h_0}{d} \Bigr)^3 \Rightarrow \mu \sim
    \Bigl(\frac{D^3}{L^2d} \Bigr)^{1/3} S^{2/3}\,.
    \label{eq:scaling}
\end{equation}
Thus, the prediction of anomalous scaling at large $S$ is robust, and should hold for both the long- and short-bearing limits in conformal contacts with an $h_0$-independent step height $d$. 

\begin{figure}
  \centering
  \includegraphics[width=\columnwidth]{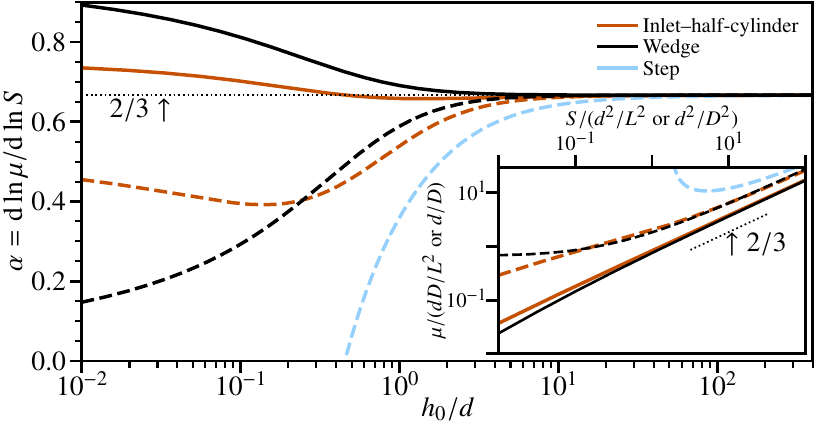}
  \caption{
    Running exponent of Stribeck curve,
    $\alpha = \leibnizd{\,\ln\mu}/\leibnizd{\,\ln S}$,
    against gap, $h_0/d$, from
    Reynolds lubrication theory,
    for various conformal profiles (legend)~\cite{PRE}. Lines: solid, short-bearing; dashed, long bearings (relative inlet length = 0.5). Short-bearing profiles: wedge, $\langle \delta h \rangle /d = 0.5$; `blunt' inlet--half-cylinder with relative inlet length such that $\langle \delta h \rangle /d = 0.4$. Inset: corresponding $\mu(S)$.}
  \label{fig:stribeck}
\end{figure}

For $S\lesssim S^*$ the gap shrinks to $d \lesssim h_0$ and the above scaling arguments no longer apply; rather, the actual gap profile, $h(x)$, must be used to calculate  $\mu(S)$ parametrically for different profiles (\partFig{fig:ball}{d}) and bearing types, \Fig{fig:stribeck}~(inset) \cite{PRE}. Deviations can be highlighted by reporting the `running exponent' $\alpha = \leibnizd{\,\ln\mu}/\leibnizd{\,\ln S}$ as a function of $h_0/d$~\cite{PRE}, \Fig{fig:stribeck}.  Asymptotically, all profiles collapse to $\alpha={2}/{3}+\order{d/h_0}$ verifying $S^{2/3}$ scaling for short and long slider bearings. For $h_0\alt d$, $\alpha$ deviates from this large-gap scaling, with the leading order correction depending on moments of the height profile~\cite{PRE}.  Typically, the Stribeck curve deviates positively ($\alpha < 2/3$) as $h_0\to d$\,; this is the case for most long-limit bearings, \Fig{fig:stribeck}~(dashed lines), and for surface profiles that are `blunt' in the sense that $\langle \delta h\rangle/d<1/2$, where $\delta h = h - h_0$. Thus, $\mu(S)$ resembles the onset of EHL, but the physics arises entirely from HL with two independent length scales.   In the limit where the gap shrinks to zero the behaviour is set by the type of profile.  For example the inlet--half-cylinder notably segues into a cylinder-on-flat geometry as $h_0\to 0$, with $\alpha = 1/2$. For other details, see~\Refcite{PRE}.

\begin{figure}[t]
  \includegraphics[width=\columnwidth]{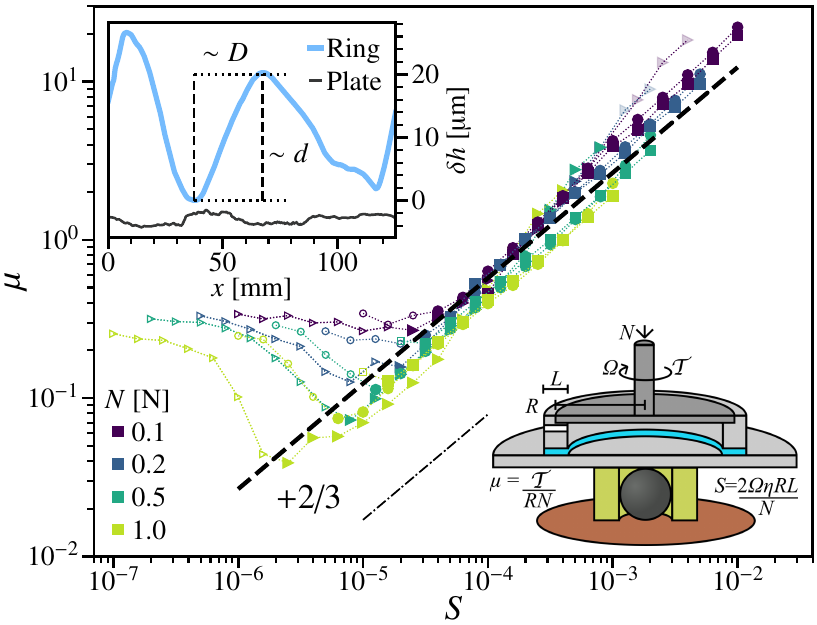}
  \caption{Self-aligning ring-plate tribo-rheology. Stribeck curves, $\mu(S)$ at different loads, $N$ (see legend), and viscosities [$\eta = 5$ (triangles), 50 (squares) or \SI{500}{\milli\pascal\second} (circles)] with $\mu = \mathcal{T}/RN$ and $S = 2\Omega\eta RL/N$. Lines, $\mu \sim S^{2/3}$: bold dashed, data fit; fine dot-dashed, scaling with unity pre-factor. Upper inset: profile, $\delta h(x)$ for plate (fine) and ring (bold) with bearing dimensions $D$ and $d$. Lower inset:~geometry cross-section with light grey, aluminium; dark grey, steel; and, yellow, foam. Ring width, $L$; radius, $R$; rotation rate, $\Omega$; and torque $\mathcal{T}$.}
  \label{fig:triborheo}
\end{figure}

Experimental verification of these predictions requires bespoke measurements, as the overwhelming majority of literature data pertains to non-conformal geometries in the EHL-BL regime.  We modified a commercial rheometer (Kinexus Ultra+, Malvern Instruments) to incorporate a ring-plate geometry (\Fig{fig:triborheo}, lower inset), inner and outer radii ($R_i$, $R_o$) respectively 17.5 and \SI{22.5}{\milli\metre}~\cite{kavehpour2004tribo}, giving $L = \SI{5}{\milli\meter}$.  The ring can be considered a narrow slider bearing [$L\ll 2\pi R = \pi(R_o +R_i)$] wrapped around upon itself. A $\mu\sim S^{2/3}$ regime has previously been observed for a ring-plate geometry and interpreted in terms of geometry misalignment, where non-parallelism creates an effective wedge angle~\cite{clasen2013high}. However, such misalignment creates an ill-defined, rotation-dependent gap profile. For a consistent gap profile, we use a self-aligning mechanism adapted from \Refcite{clasen2013self}.  A flexible foam mounting allows the plate to tilt about a central ball bearing, but not freely rotate, and the applied load dynamically pushes the shearing surfaces parallel. Surfaces are used as machined. 

To measure gap profiles we rigidly mount the plate or ring as the lower geometry and attach a \SI{10}{\milli\metre} polytetrafluoroethylene-coated sphere  centred at \SI{20}{\milli\metre} from the upper geometry rotated at $\Omega = \SI{0.1}{\radian\per\second}$ while imposing a \SI{0.02}{\newton} normal force through a feedback loop. The change in gap needed to maintain contact over a cycle to first order gives the tilt of the rigid mounting, which is compensated for in the self-aligning geometry. Subtracting the tilt leaves  
the bearing surface profile, $\delta h$ (\Fig{fig:triborheo}, upper inset). The plate is flat on the $\sim \SI{1}{\micro\meter}$ level, but the rixng has undulations of $d\approx \SI{22}{\micro\metre}$ and $D \sim \pi R/2 \approx \SI{30}{\milli\metre}$ acting as two symmetric wedge bearings in series in which we set $p=0$ in the two divergent halves by appealing to the half-Sommerfeld boundary condition~\cite{gropper2016hydrodynamic} (\cf\ Fig.~3(a) in \Refcite{PRE}). In such a short bearing, for which Eq.~\ref{eq:scaling} applies, lubricant leaks from the sides; to prevent bearing starvation, excess fluid from loading is left as a reservoir~\cite{Note1}.

Three poly(dimethyl siloxane) silicone oils (Merck, UK) were used ($\eta = 5$, 50 and \SI{500}{\milli\pascal\second})~\cite{SM}. We controlled the initial temperature of the sample using a Peltier plate at $T=\SI{20}{\celsius}$. The maximum temperature rise during measurements due to viscous heating is $\lesssim \SI{2}{\celsius}$~\cite{SM}; $\leibnizd\eta/\leibnizd T$ data on silicone oils~\cite{Roberts2017} suggest that this has negligible effect for our work.
The load was varied ($N = 0.1$ to \SI{1.0}{\newton}) for logarithmic sweeps of the rotation rate, $\Omega$, from \SI{0.1}{\radian\per\second} upwards at 5~pts/decade, until reaching a maximum torque (\SI{0.05}{\newton\metre}) or sample ejection (at $\Omega_{\max} \approx \SI{150}{\radian\per\second}$). To average over multiple rotations, the step time was \SI{100}{\second} for $\Omega < \SI{1.0}{\radian\per\second}$, and \SI{20}{\second} above, leaving \SI{10}{\second} to reach a steady state. From the torque, $\mathcal{T}$, $\mu = \mathcal{T}R/(R_o^2+R_i^2)$ [$\approx \mathcal{T}/RN$ for $L\ll R$]. In this context $S$ (or G\"umbel number) $= 2\eta \Omega R L/N$, featuring the linear speed of the bearing ($\Omega R$) and a factor of two from the ring undulations forming \emph{two} slider bearings. 

At $S \geq 6 \times 10^{-5}$, \Fig{fig:triborheo}~(bold), the majority of our data collapse  with $\load$ (increasing, dark to light) and $\eta$ (symbols). Power law fits of $\mu \sim S^{\alpha}$ for $\eta_s = 50$ and \SI{500}{\milli\pascal\second} give $\alpha = 0.72 \pm 0.05$, close to the predicted $2/3$ scaling for 
the large-gap lubrication regime (bold dashed line). Further, using the measured $d$ the predicted $\mu(S)$ is within a factor of $\approx 7$ (fine dot-dashed line), consistent with a scaling argument neglecting $\mathcal{O}(1)$ prefactors. Selected runs were also performed with $\Omega \to -\Omega$ and gave similar results~\cite{SM}, consistent with the near-symmetrical surface profile, \Fig{fig:triborheo} (upper inset). 

At $S \gtrsim 10^{-3}$ and the lowest viscosity ($\eta = \SI{5}{\milli\pascal\second}$), the curve steepens (greyed symbols). In this regime, fluid inertia becomes important: the predicted $h_0 \gtrsim \SI{80}{\micro\metre}$  
[\Eq{eq:scaling}] 
with $\Omega \gtrsim \SI{100}{\radian\per\second}$ give a Reynolds number $\mathrm{Re} = \rho \Omega R h_0/\eta \gtrsim 30$, where secondary flows~\cite{Macosko1994} and other complications arise. 

There are also deviations from $2/3$ scaling at $S \lesssim 6 \times 10^{-5}$, \Fig{fig:triborheo}~(open symbols). The data no longer collapse when plotted against $S(\load,\eta)$, but depend on $\load$ and $\eta$ separately. As $S \to 0$, $\mu$ converges to $\approx 0.3$, consistent with BL for aluminium-aluminium contact~\cite{edge}. 
Between $\mu \approx 0.3$ and $\mu \sim S^{2/3}$,  the behaviour appears similar to EHL~\cite{stachowiak2013engineering}. However, for $\load = \SI{0.1}{\newton}$ [dark (purple)] the deviation point, $S^{*} = 6\times10^{-5} \sim (d/L)^2(h_0/d)^3$, corresponds to $h_0 \approx d \sim \SI{20}{\micro\meter}$, far above the scale of asperities whose deformation triggers the onset of EHL. On the other hand, $h_0 \approx d$ is where we predict the onset of {\it hydrodynamic} deviation from $\mu \sim S^{2/3}$ scaling, \Fig{fig:stribeck}. Our geometry has a calculated average surface profile of $\langle \delta h \rangle /d = 0.41 < 1/2$, \ie\ just `blunt' enough for us to expect weak positive deviations in the Stribeck curve ($\alpha < 2/3$ as $h_0 \to d$). (Compare the solid dark orange curve in \Fig{fig:stribeck} (inset) calculated for an inlet--half-cylinder with $\langle \delta h \rangle/d = 0.4$~\cite{PRE}.) This is not the form of deviations we observe. One possible reason is that deformations in our geometry lead to a load-dependent $d(N)$, although measurements of the axial compliance of our rheometer~\cite{SM} reveal no such deformations.

More interestingly, if our interpretation of the $S^{2/3}$ scaling in terms of an $h_0$-independent step height $d$ is correct, then as $h_0\to d$, multiple other length scales should become relevant and change the functional form of $\mu(S)$. Likely candidates include  mesoscale roughness in the plate (\Fig{fig:triborheo}, upper inset, black line; see also~\Refcite{SM}) or the ring. It is only if highly polished components are used that such `extra length scales' will disappear and allow short-wedge-like deviations from $S^{2/3}$ scaling to show through. Instead, the form of $\mu(S)$ we obtain using `as-machined' components, \Fig{fig:triborheo}, probably represents the most likely encountered generic case. 

There are few modern Stribeck curve data for the HL regime extensive enough to test for scaling. Recently, Classen has twice reported $\mu \sim S^{2/3}$ at high $S$, but interpreted this, and earlier data~\cite{kavehpour2004tribo}, in terms of an effective geometry misalignment~\cite{clasen2013high,clasen2013self}. Studies in which there has also been independent measurement of the surface profile are even rarer. One exception is the work already mentioned  on molten chocolate in conditions corresponding to oral processing \cite{soltanahmadi_insights_2023}.  This compares a single lightly loaded smooth elastomeric ball on flat glass to a comparably loaded bio-mimetic surface with multiple \emph{rough} contact points (predominantly flat-topped cylinders). At large gaps and sliding speeds the molten chocolate can be considered a Newtonian fluid with $\eta \approx \SI{1}{\pascal \second}$.

The high-$S$ scaling behaviour of the two surfaces is notably different  (\Fig{fig:tongue}). The data for a ball-on-flat geometry increases from a plateau ($S \lesssim 10^{-3}$) with a steepening gradient. At $S\gtrsim 10^{-2}$ the trend reaches $\mu \sim S^\alpha$ with $\alpha=0.8\pm0.1$ from linear regression, and the data plausibly tends towards a linear dependence (bold dashed line). In contrast, for a conformal textured surface in contact over a large area ($2\times2\,\si{\centi\metre^2}$), for $S> 4 \times 10^{-3}$ we find a similar power law but with an exponent $\alpha=0.6\pm0.1$ and no sign of tending to linear scaling (light dotted line) over 1.5 decades.  Instead, the data appear consistent with our predicted $\mu \sim S^{2/3}$ for two competing length scales (bold dotted line). In \Refcite{PRE}, we analyse this data in detail as HL between a smooth steel plate and individual `papillae' on the biomimetic tongue that are step-textured on the $d \sim \SI{50}{\micro\meter}$ scale, which we can deduce from the point at which deviations from $S^{2/3}$ scaling is first observed.

To summarise, revisiting the classic HL regime for conformal contacts reveals a Stribeck curve distinct from that expected for non-conformal contacts. In particular, our analysis and experiments support $\mu \sim S^{2/3}$ in the high-$S$ limit, wherein the exponent is not set by dimensionality, but signals the presence of two independent length scales, the bearing length and a step height.  This anomalous scaling applies in the large-gap limit, where the gap is greater than the step height. When these become comparable at low enough $S$, deviations from $\mu\sim S^{2/3}$ are expected. We tested these predictions using tribo-rheology, with a novel combination of surface profile characterisation and a bespoke self-aligning geometry. The results were consistent with our HL scaling analysis at large $S$. At small $S$ the results indicate the presence of additional length scales for surfaces `as machined'. Comparison with  literature data under lightly-loaded conditions relevant to oral processing~\cite{soltanahmadi_insights_2023} provides further experimental support for our contention that the high-$S$ HL behaviour of non-conformal contacts with a single length scale, $\mu \sim S$, differs fundamentally from that of conformal contacts with competing length scales, $\mu \sim S^{2/3}$.   

Beyond intrinsic interest, the subtleties of the HL regime in flat-flat contacts that we have uncovered may have particular relevance for sensory physics. The application of topical cosmetics and medicines involves traversing the entire Stribeck curve from high to low $S$ with the product as the lubricant, starting with a low load and large gap~\cite{adams2007friction}. The same considerations also apply to the oral perception of many foods~\cite{Pradal2016}. In all these cases, the two length scales traceable to machining in our experimental geometry are also likely present, but as surface texturing or roughness. The generic features of the Stribeck curve in \Fig{fig:triborheo} should therefore recur in these and other areas of applications involving human texture perception~\cite{VanAken2010,Zhang2017}. 

\newcommand{\PBW}{P.\,B.\,W.}
\newcommand{\WCKP}{W.\,C.\,K.\,P.}
\newcommand{\JAR}{J.\,A.\,R.}

\acknowledgments{\PBW, \WCKP\ and \JAR\ conceptualised the work and drafted the manuscript. Experiments were carried out by \JAR, and calculations by \PBW\ and \WCKP; all authors interpreted data and revised the manuscript. We thank Rory O'Neill for technical assistance and Andreia Silva and John Royer for useful discussions.}


%

\end{document}